\documentclass[10pt, letterpaper, conference]{IEEEtran}

\IEEEoverridecommandlockouts

\usepackage{graphicx}
\usepackage{subfigure}
\usepackage{curves}
\usepackage{amsfonts}
\usepackage[cmex10]{amsmath}
\usepackage{epsfig}
\usepackage{stfloats}
\usepackage{amssymb}
\usepackage{cite}
\usepackage{epstopdf}
\usepackage{bm}
\usepackage{bbm}

\usepackage[vlined,linesnumbered,ruled]{algorithm2e}

\SetFuncSty{functionfont}
\SetCommentSty{commentfont}
\SetAlFnt{\small}

\begin{document}

\title{PAC Codes for Source and Joint \\Source-Channel Coding}

\author{%
	\IEEEauthorblockN{Mengfan Zheng\IEEEauthorrefmark{1}
		and Cong Ling\IEEEauthorrefmark{2}}
	\IEEEauthorblockA{\IEEEauthorrefmark{1}%
		Department of Electronic and Computer Engineering, The Hong Kong University of Science and Technology}
	\IEEEauthorblockA{\IEEEauthorrefmark{2}%
		Department of Electrical and Electronic Engineering, Imperial College London}
	Emails:  eemzheng@ust.hk, 
	c.ling@imperial.ac.uk
}
\maketitle

\begin{abstract}
	Polarization-adjusted convolutional (PAC) codes, as a concatenated coding scheme based on polar codes, is able to approach the finite-length bound of binary-input AWGN channel at short blocklengths. In this paper, we extend PAC codes to the fields of source coding and joint source-channel coding and show that they can also approach the corresponding finite-length bounds at short blocklengths.
\end{abstract}

\section{Introduction}
Polarization-adjusted convolutional (PAC) codes, proposed by Ar{\i}kan in 2019 \cite{arikan2019sequential}, is a concatenation scheme of polar codes which greatly improves the finite-length performance. The idea is to apply a convolutional pre-transform on the uncoded bits before feeding them to the polarized synthetic channels. It is shown that using a proper convolutional transform and a powerful decoder, such a scheme can approach the finite-length bound (or dispersion bound) of the binary-input AWGN channel at rate $R=0.5$ and code lengths $N=128$ and $N=256$ \cite{arikan2019sequential,rowshan2020polarization,yao2020list,tonnellier2020systematic}. 

PAC codes were originally proposed as a channel coding scheme. It has been shown that polar codes also have very good finite-length performance in lossless source coding \cite{Zheng2021jscc} and joint source-channel coding (JSCC) \cite{Zheng2021jscc,Dong2021JSCC,Dong2022JSCC}. Although for the lossless source coding problem, our previous work has shown that polar codes with CRC-aided successive cancellation list (CA-SCL) decoding can approach the finite-length bound, for the JSCC problem, as far as we know, there has not been any finite-length-bound-approaching scheme in the literature yet. This makes us wonder if we can use PAC codes to achieve this goal.

The motivation for studying JSCC is that  Shannon’s source–channel separation theorem \cite{shannon1948mathematical}, which states that we can gain nothing from the joint design of source and channel coding asymptotically, only holds when the delay is unbounded. In the finite blocklength regime, JSCC is strictly better than separate source-channel coding (SSCC). When the blocklength is small, source coding will have a certain amount of residual redundancy. If the residual redundancy can be exploited by the channel decoder properly, the overall error performance could be improved. 

The have been some researches on polar code-based JSCC schemes for better finite-length performance. A joint decoder that combines SCL polar decoder and language decoder is proposed in \cite{wang2016lang}, which exploits the redundancy of language-based sources during polar decoding to improve error performance. In \cite{wang2017source}, it is shown that the rate of polar codes can be improved by exploiting source redundancy. In that work, source redundancy is simply modelled using a sequence of $t$-erasure correcting block codes, which lacks generality. In \cite{Zheng2021jscc}, we propose a joint source-channel polar coding scheme and a powerful joint decoding scheme which can break through the finite-length bound of SSCC at short blocklengths. In \cite{Dong2021JSCC,Dong2022JSCC}, a similar scheme with different decoding algorithms are proposed.

In this paper, we further improve our previous work \cite{Zheng2021jscc} by adopting the PAC codes as both the source and the channel component codes. First, we extend PAC codes to source coding and show that they also outperform polar codes. Then we combine source and channel PAC codes in the JSCC scenario and propose a joint source-channel PAC coding scheme, which is shown to approach the finite-length bound of JSCC over the binary-input AWGN (BI-AWGN) channel at short blocklengths. To the best of our knowledge, this is the first scheme that can achieve this so far.

\textit{Notations:} $[N]$ is the abbreviation of an index set $\{1,2,...,N\}$. Vectors and matrices are denoted by boldface letters. Vectors are also denoted as $X^{a:b}\triangleq \{X_a,X_{a+1},...,X_{b}\}$ for $a\leq b$. For a subset $\mathcal{A}\subset [N]$, $X^{\mathcal{A}}$ denotes the subvector $\{X_i:i\in\mathcal{A}\}$ of $X^{1:N}$. $\mathcal{A}^C$ ($\mathcal{A}\subset [N]$) denotes the complementary set of $\mathcal{A}$ in $[N]$. $\mathbf{G}_N=\mathbf{B}_N \textbf{F}^{\otimes n}$ is the generator matrix of polar codes \cite{arikan2009channel}, where $N=2^n$ with $n$ being an arbitrary integer, $\mathbf{B}_N$ the bit-reversal matrix, and $\textbf{F}=
\begin{bmatrix}
	1 & 0 \\
	1 & 1
\end{bmatrix}$. $\delta_N=2^{-N^\beta}$ with some $\beta \in (0,1/2)$.

\section{Preliminaries on Polar and PAC Codes}

\subsection{Polar Codes}
Polar codes are defined by the polar transform \cite{arikan2009channel}:
\begin{equation}
	\mathbf{x}=\mathbf{u}\mathbf{G}_N, \label{encoding}
\end{equation}
where $\mathbf{u}=[u_1,...,u_N]$ is the uncoded bit sequence and $\mathbf{x}=[x_1,...,x_N]$ is the encoded codeword. The construction problem of polar codes is to partition $\mathbf{u}$ into an \textit{information set} $\mathcal{I}$ and a \textit{frozen set} $\mathcal{F}=\mathcal{I}^c$. Message bits are assigned to $\mathbf{u}^{\mathcal{I}}$ while $\mathbf{u}^{\mathcal{F}}$ are assigned with some fixed value, such as 0. 

Upon receiving $\mathbf{y}$, the receiver can use a successive cancellation (SC) decoder to recover $\mathbf{u}$:
\begin{equation*}
	\hat{u}_{i}=
	\begin{cases}
		u_i,&\text{ if } i\in \mathcal{F}\\
		\arg\max_{u\in\{0,1\}}P_{U_{i}|\mathbf{Y}, U^{1:{i-1}}}(u|\mathbf{y},\hat{u}^{1:{i-1}}),&
		\text{ if } i\in \mathcal{I}
	\end{cases}.
\end{equation*}

To improve finite-length performance, a common way is to use the CA-SCL decoding \cite{niu2012CRC,Tal2015List}. The idea is to retain up to $L$ most probable paths during the SC decoding process and use CRC to select the correct one.

\subsection{PAC Codes}

\begin{figure}[t]
	\centering
	\includegraphics[width=0.88\columnwidth]{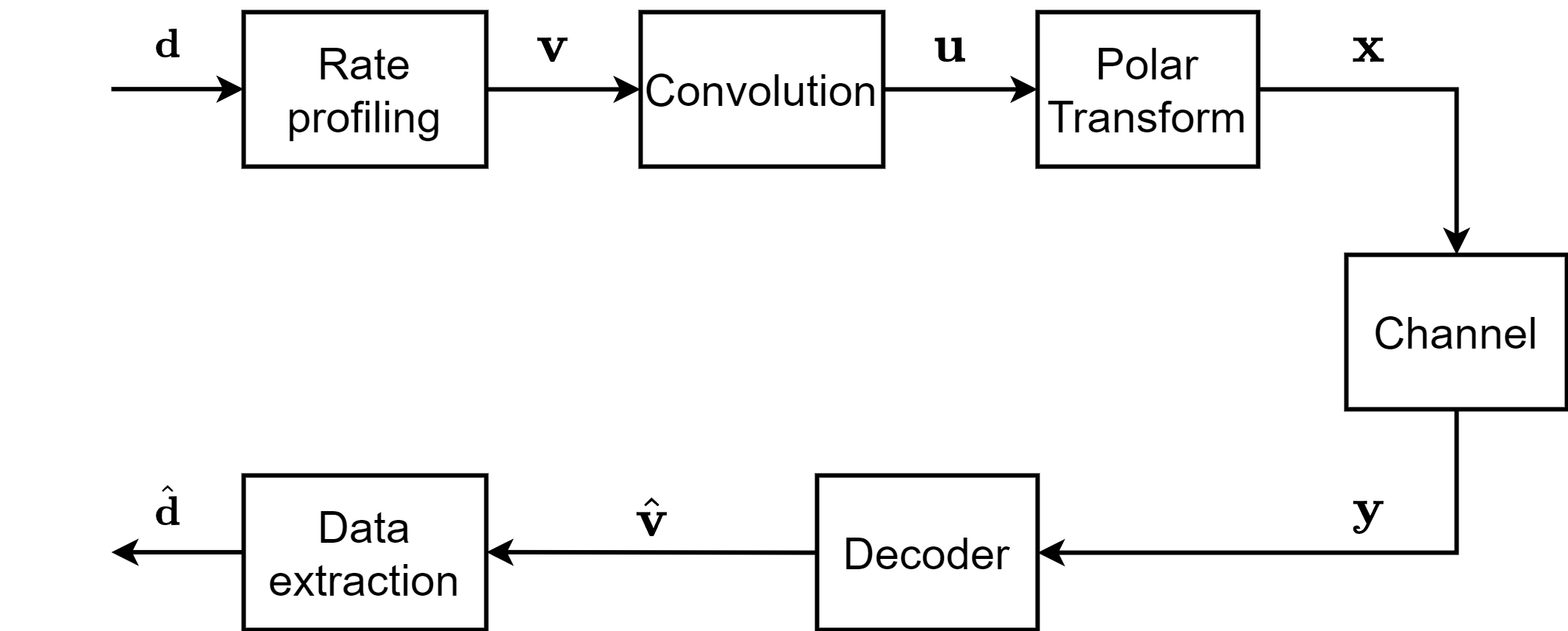}
	\caption{PAC coding scheme \cite{arikan2019sequential}.} 
	\label{fig:PAC}
\end{figure}

Fig. \ref{fig:PAC} illustrates the PAC scheme proposed by Ar{\i}kan. A rate profiler first maps the information bits $\mathbf{d}$ to an $N$-bit vector $\mathbf{v}$. Then, the convolutional transform with polynomial coefficients vector $\mathbf{g}$ scrambles $\mathbf{v}$ and outputs $\mathbf{u}$. $\mathbf{u}$ is then fed to the polar transform. The convolutional transform creates correlation between bits sent to different synthetic sub-channels, which can be exploited by the decoder. The rate profile (i.e., how $\mathbf{d}$ is mapped to $\mathbf{v}$) determines the construction of a PAC code. Just like polar codes, the performance of a PAC code greatly depends on the code construction. Ar{\i}kan shows in \cite{arikan2019sequential} that for $N=128$ and $R=1/2$, the Reed-Muller (RM) rate profile (which is used to generate Reed-Muller codes) seems to provide the best performance. However, for other code lengths and rates, this may not hold.

Suppose the polynomial coefficients vector of the convolutional transform is $\mathbf{g}=[c_0  c_1  c_2   \cdots c_\nu]$. Then the convolutional transform can be represented by the upper-triangular Toeplitz matrix:		
\begin{equation}
	\mathbf{T}=
	\begin{bmatrix}
		\,c_0  & c_1    & c_2    & \cdots & c_\nu   & 0      & \cdots & 0    
		\\[-1.26ex]
		0      & c_0    & c_1    & c_2    & \cdots & c_\nu   &        & \vdots 
		\\[-1.26ex]
		0      & 0      & c_0    & c_1    & \ddots & \cdots & c_\nu  & \vdots 
		\\[-1.26ex]
		\vdots & 0      & \ddots & \ddots & \ddots & \ddots &       & \vdots 
		\\[-1.26ex]
		\vdots & \phantom{\ddots}& \ddots & \ddots & \ddots & \ddots & c_2     & \vdots 
		\\[-1.26ex]
		\vdots &        &        & \ddots & 0      & c_0    & c_1   & c_2    
		\\[-1.26ex]
		\vdots &        &        &        & 0      & 0      & c_0   & c_1    
		\\[-1.26ex]
		0      & \cdots & \cdots & \cdots & \cdots & 0      & 0     & \,c_0\phantom{\vdots}    
	\end{bmatrix}
\end{equation}
Thus, the encoding of a PAC code can be written as: 
\begin{equation}
	\mathbf{x}=\mathbf{v}\mathbf{T}\mathbf{G}_N.
\end{equation}

For the decoding part, Ar{\i}kan proposes to use sequential decoding, which goes as follows. The decoder tries to identify the correct path in the code tree by using a metric that tends to increase along the correct path and decrease as soon as a path diverges from the correct path. When the path metric falls below a threshold, the decoder backtracks and chooses another path. When there is no path above the threshold, the threshold is decreased. The path metric can be computed using a low-complexity recursive method, as in SC decoding of polar codes. However, the biggest issue of this decoder is that the complexity is not fixed since we do not know how many times the decoder will trackback. For various SNRs, the average decoding latency of sequential decoding can be orders of magnitude different. To overcome this problem, list decoding has been studied \cite{rowshan2020polarization,yao2020list}, which is shown to be able to achieve comparable performance to sequential decoding. List decoding of PAC codes is similar to that of conventional polar codes, except that when decoding an information bit, the influence of its previous bits need to be subtracted.

\section{Source PAC Coding}
\label{Sec:SourcePAC}

\subsection{Source Polar Code}
Consider the almost-lossless compression\footnote{Since we are considering fixed-to-fixed length compression, lossless recovery is achieved as the blocklength goes to infinity. At finite blocklengths, there will always be decoding errors.} of a memoryless binary source $X\sim p_{X}$. Let 
\begin{equation}
	U^{1:N}=X^{1:N}\mathbf{G}_N.
\end{equation} 
As $N$ goes to infinity, $U^j$ ($j\in [N]$) polarizes in the sense that it is either almost independent of $(U^{1:j-1})$ and uniformly distributed, or almost determined by $(U^{1:j-1})$ \cite{arikan2010source}. Based on this, we can define the following low-entropy set of the polarized indices:
\begin{align}
	\mathcal{L}^{(N)}_{X}=\{j\in [N]:H(U^j|U^{1:j-1})\leq \delta_N\},\label{LX}
\end{align}
which is proven to satisfy
\begin{align}
	\lim_{N\rightarrow \infty}\frac{1}{N}|\mathcal{L}^{(N)}_{X}|=1-H(X). \label{PolarRate2}
\end{align}

Define $\mathcal{H}^{(N)}_{X}=(\mathcal{L}^{(N)}_{X})^C$. To compress the source sequence $X^{1:N}$, we only need to retain $U^{\mathcal{H}^{(N)}_{X}}$ and abandon the rest, since $U^{\mathcal{L}^{(N)}_{X}}$ can be determined successively with high probability given the value of $U^{\mathcal{H}^{(N)}_{X}}$.

\subsection{Source PAC Code Design}

\begin{figure}[t]
	\centering
	\includegraphics[width=0.87\columnwidth]{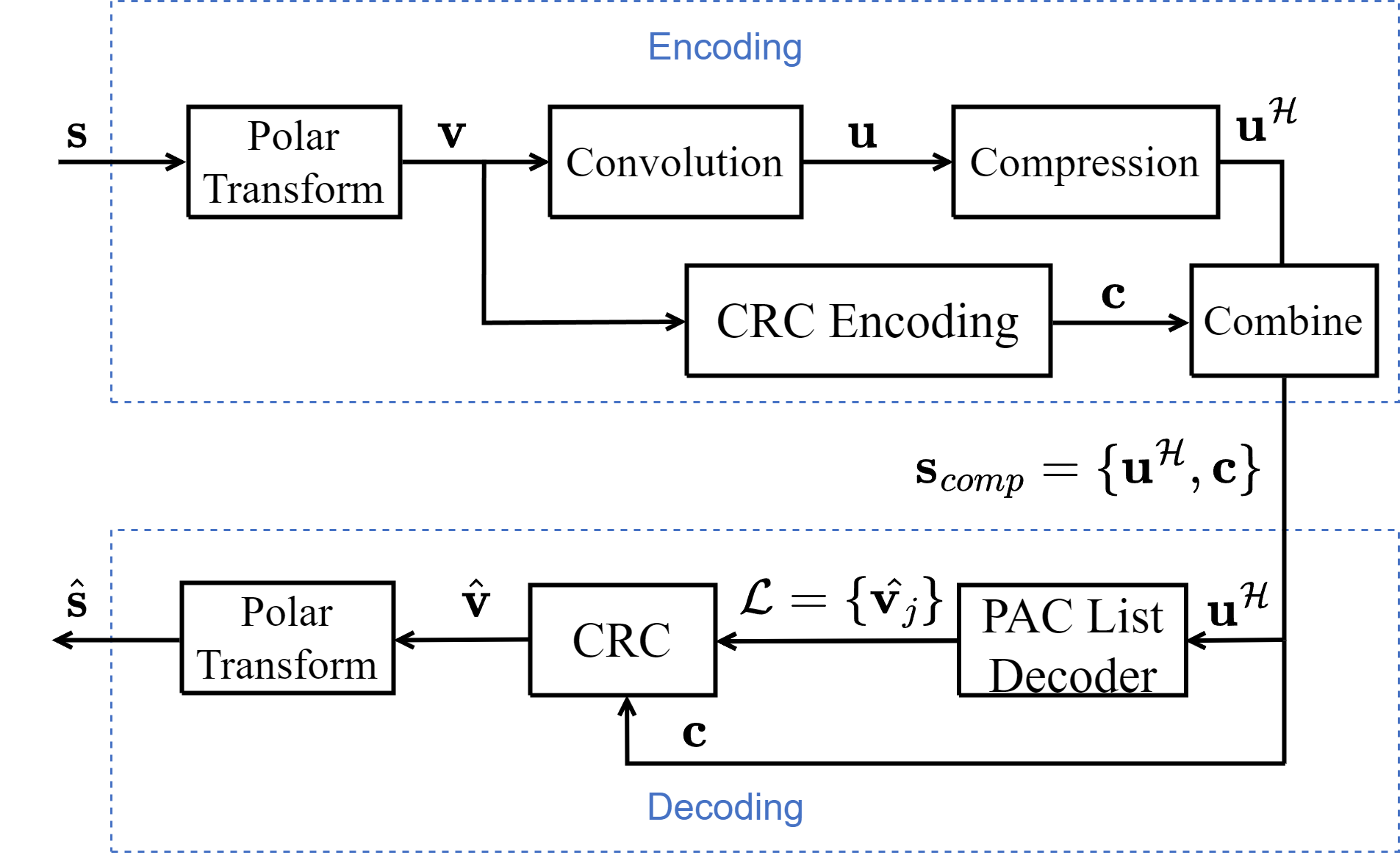}
	\caption{CRC-aided source PAC coding scheme.} 
	\label{fig:SourcePAC}
\end{figure}

Inspired by the idea of PAC codes for channel coding, we extend this approach to source coding. Just like source polar codes, CRC can also be used in source PAC codes to improve finite-length performance. Fig. \ref{fig:SourcePAC} illustrates the block diagram of our proposed CRC-aided PAC (CA-PAC) source coding scheme, where $\mathbf{s}$ is the length-$N$ source sequence to be compressed, and $\mathbf{s}_{comp}$ is the encoding output. In contrast to PAC channel coding, for source coding the convolutional transform is applied to the polar-transformed sequence $\mathbf{v}=\mathbf{s}\mathbf{G}_N$, which generates a vector $\mathbf{u}$, where
\begin{equation}
	\mathbf{u}=\mathbf{s}\mathbf{G}_N\mathbf{T}.
\end{equation}
Then only a fraction of $\mathbf{u}$ (denoted by $\mathcal{H}$, referred to as the high-entropy set) is retained. Finally, CRC bits of $\mathbf{v}$, denoted by $\mathbf{c}$, are appended to $\mathbf{u}^{\mathcal{H}}$, which completes the encoding process. Therefore, the encoding result can be written as
\begin{equation}
	\mathbf{s}_{comp}=\{\mathbf{u}^{\mathcal{H}}, \mathbf{c}\}.
\end{equation}

The choice of $\mathcal{H}$ determines the construction of a PAC source code, similar to the rate profile in PAC channel codes. By simulations we found that the RM rule does not work well for PAC source codes. Thus, $\mathcal{H}$ is chosen to be the same as that for conventional source polar coding in our simulations. Whether there exists better choices of $\mathcal{H}$ is left for future research.

To recover $\mathbf{s}$ from $\mathbf{s}_{comp}$, we first use a PAC list decoder to obtain a list of candidate estimates for $\mathbf{v}$, denoted by $\mathcal{L}=\{\mathbf{\hat{v}}_i\}$ and sorted according to their probability. Then CRC is performed on $\mathbf{\hat{v}}_i$ to select the best estimate of $\mathbf{\hat{v}}$. Finally $\mathbf{\hat{v}}$ is inverse polar-transformed to retrieve the original source. 

\subsection{Performance}
\label{SubSec:SCLSource}

\begin{figure}[t]
	\centering 
	\subfigure[N=128]{ 
		\includegraphics[width=0.98\columnwidth]{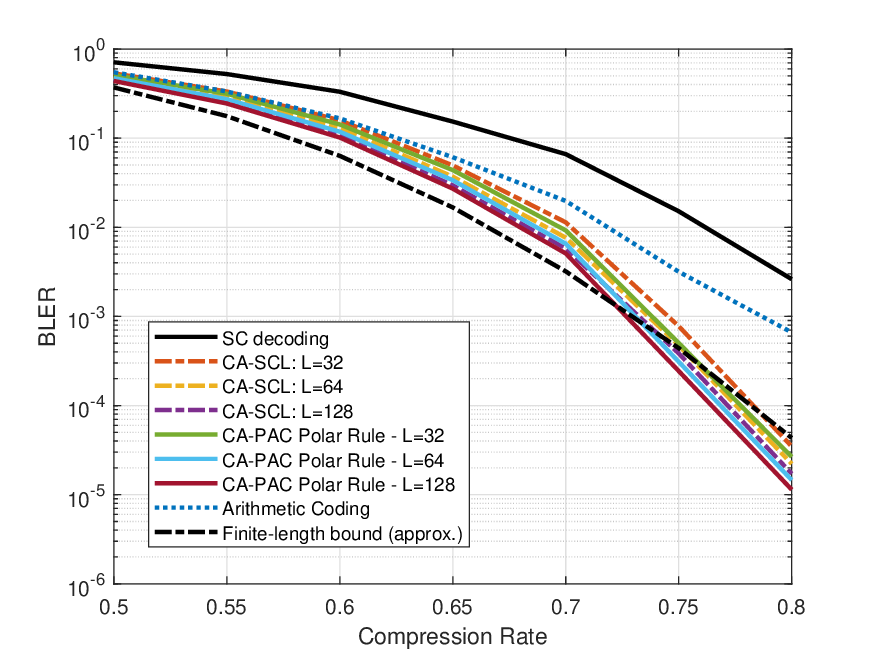}} 
	\subfigure[N=256]{  
		\includegraphics[width=0.98\columnwidth]{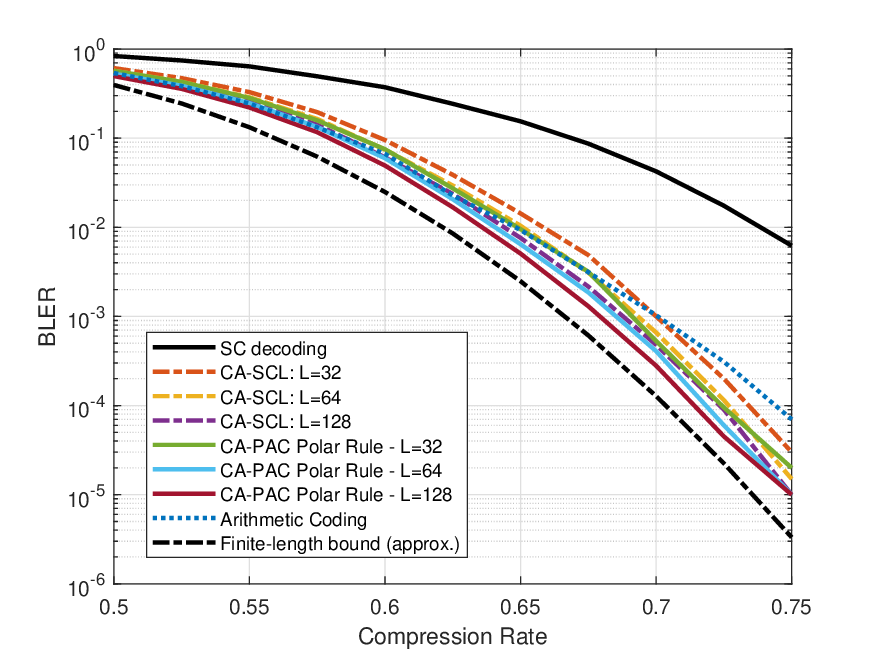}}
	\caption{Performance of source PAC codes. $P(0)=0.89$, $N=128, 256$, $CRC=8$.} 
	\label{fig:PACSource}
\end{figure}

Now we compare the performance of the proposed CA-PAC codes with the CRC-aided source polar codes in \cite{Zheng2021jscc}. We consider compressing a biased Bernoulli source with $\mathrm{P}(1)=0.11$ (denoted as $\text{Bern}(0.11)$) at block-length $N=128$ and $256$ . For the CA-PAC scheme, we choose $\mathbf{g}=[1 1 0 1 0 1 1 0 1 0 1 1]$. The CRC length is 8 for both schemes. In addition, we also compare with arithmetic coding in this example, since arithmetic coding can produce near-optimal output for any given set of symbols and probabilities. Because the output length of arithmetic coding is not fixed, for fare comparison, we truncate over-length codewords to the same fixed length in simulations. The results are shown in Fig. \ref{fig:PACSource}. It can be seen that the CA-PAC scheme outperforms the CA-SCL scheme of the same list size, and both of them outperforms arithmetic coding for $N=128$ and $N=256$. Also, in both cases, the finite-length bounds of fixed-length source compression \cite{Kostina2012lossy} are approached with a very small gap. The reason that some schemes even outperform the bound for $N=128$ may be that the bound is calculated according to an approximation formula and is not accurate enough at very short blocklengths. This result reflects that polar-based source coding schemes have great potential in the short blocklength regime where traditional compression algorithms do not work well.

Although the performance gain of CA-PAC codes compared with CRC-aided source polar codes may not be so significant, the convolutional transform provides more opportunity for joint source-channel decoding, as we will show in the next section.

\section{Joint Source-Channel PAC Code Design}
\label{Sec:Scheme}

\subsection{JSCC PAC Code Design}

Since both source and channel PAC codes can approach the corresponding finite-length bounds at short block-lengths, it is natural to think that their concatenation can also approach the finite-length bound of SSCC. A more ambitious question is, can we design a proper joint decoding scheme for the source and channel PAC codes to approach the JSCC bound? In this section, we show that the answer is yes.

\begin{figure}[t]
	\centering
	\includegraphics[width=0.86\columnwidth]{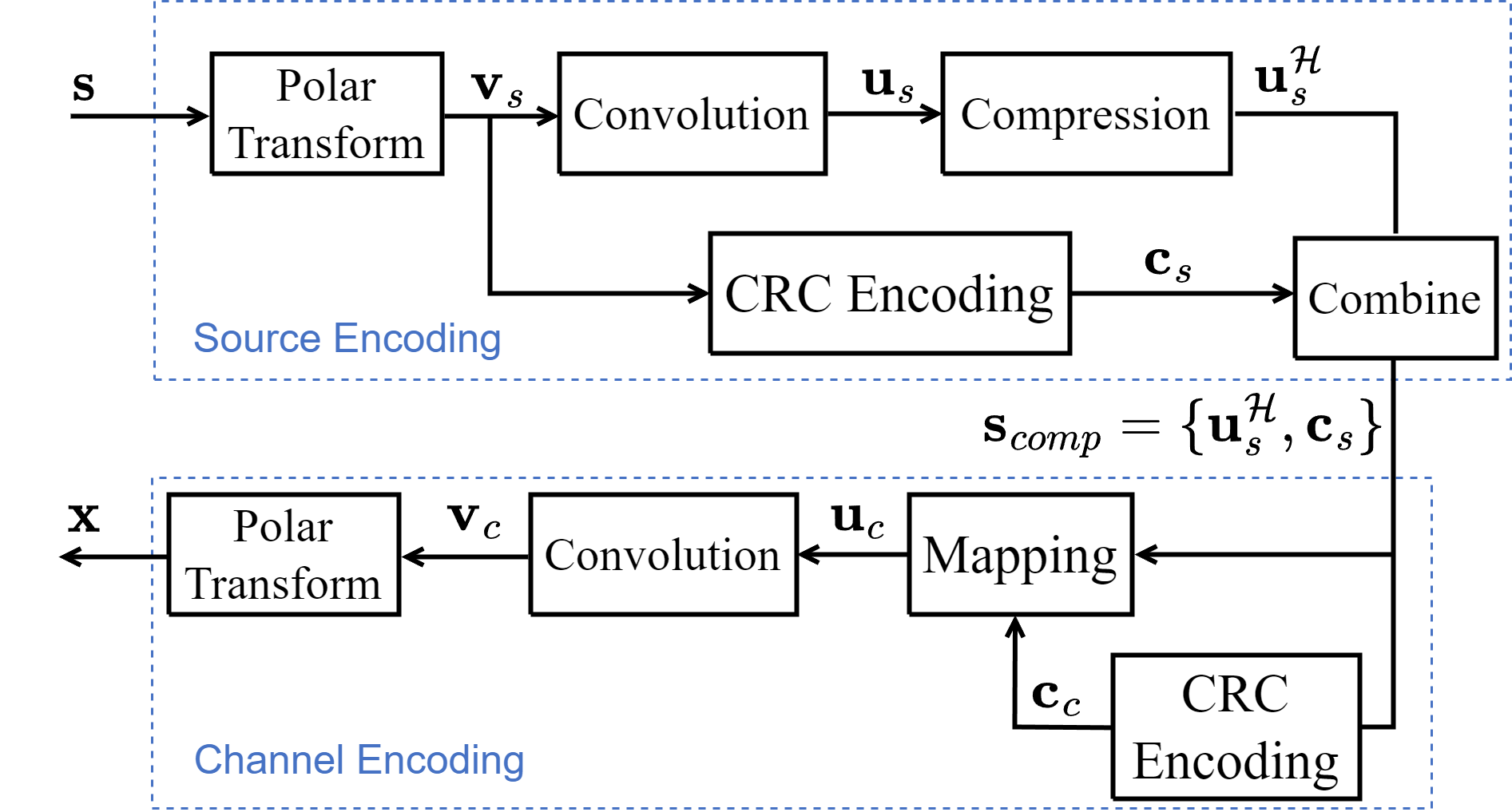}
	\caption{Joint source-channel PAC coding scheme.} 
	\label{fig:JSCCPAC}
\end{figure}

Fig. \ref{fig:JSCCPAC} shows the block diagram of our proposed joint source-channel PAC coding scheme. The encoder side is a concatenation of a source PAC code and a channel PAC code. A source sequence $\mathbf{s}$ is first compressed to $\mathbf{s}_{comp}$ using the scheme introduced in the previous section. Then $\mathbf{s}_{comp}$ is encoded using a PAC channel code. The intermediate variables in this encoding scheme are shown in Fig. \ref{fig:JSCCPAC}. In the rest of this paper, we use $\mathcal{H}$ to denote the high-entropy set in $\mathbf{u}_s$ and $\mathcal{I}$ the information bit set in $\mathbf{u}_c$. We assume that CRC bits are placed at the end, thus 
\begin{equation}
    \mathbf{u}_c^{\mathcal{I}}=\{\mathbf{u}_s^{\mathcal{H}}, \mathbf{c}_s, \mathbf{c}_c\}.\label{mapping}
\end{equation}

Note that in concatenated source-channel polar coding schemes \cite{Zheng2021jscc,Dong2021JSCC,Dong2022JSCC}, the polar-transformed source $\mathbf{v}_s$ is directly compressed into $\mathbf{v}_s^{\mathcal{H}}$, where $\mathcal{H}$ is the high-entropy set of the source polar code, and then $\mathbf{v}_s^{\mathcal{H}}$ is mapped to $\mathbf{v}_c$, the input vector of the channel polar encoder. There is no direct connection between the abandoned source bits $\mathbf{v}_s^{\mathcal{H}^c}$ and $\mathbf{v}_c$ (except for the source CRC bits if used), resulting in little opportunity for joint decoding. This may explain why the scheme in \cite{Zheng2021jscc} can only break through the SSCC bound a little at very short blocklengths ($N\leq 128$). In the concatenated PAC scheme, the convolutional transforms create more connection between $\mathbf{v}_s$ and $\mathbf{v}_c$, which can be exploited for joint decoding. 

\subsection{Joint Decoding}

The channel decoding rule used in \cite{Kostina2013JSCC} for proving the achievability of JSCC is
\begin{align}
	\hat{\mathbf{s}}_{comp}=\arg \max_{\mathbf{s}_{comp}}\mathrm{P}(\mathbf{s}_{comp}|\mathcal{C}_s)\mathrm{P}(\mathbf{y}|\mathbf{x}(\mathbf{s}_{comp})), \label{JSCCRule}
\end{align}
where $\mathcal{C}_s$ is the codebook of the source code and $\mathbf{x}(\mathbf{s}_{comp})$ is the channel encoder's output codeword given $\mathbf{s}_{comp}$ as the input. The first probability in (\ref{JSCCRule}) measures the source encoder's output distribution, while the second probability is the conditional probability of the channel. After the channel decoder determines $\hat{\mathbf{s}}_{comp}$, a source decoder then reconstruct the original source with it. 

The joint decoding algorithm used in this paper is similar to that in our previous work\cite{Zheng2021jscc} for joint source-channel polar coding. The core is to find a method to estimate the two probabilities in (\ref{JSCCRule}). In practice, it is infeasible to calculate them for every possible $\mathbf{s}_{comp}$ in order to find $\hat{\mathbf{s}}_{comp}$. Therefore, we measure the following quantity during the channel SCL decoding process\footnote{The list decoding of PAC codes is also successive cancellation in nature.} to imitate the idea behind (\ref{JSCCRule}) while limiting the search space to a manageable size:
\begin{align}
	Q(u_c^{[i]})&\triangleq \mathrm{P}(u_c^{[i]\cap \mathcal{I}}|\mathcal{C}_s)\mathrm{P}(u_c^{[i]}|\mathbf{y}) \nonumber\\
	&= \mathrm{P}(u_s^{[j]\cap \mathcal{H}}|\mathcal{C}_s)\mathrm{P}(u_c^{[i]}|\mathbf{y}),
\end{align}
where $\mathbf{u}_c$ is the input vector to the channel PAC code as shown in Fig. \ref{fig:JSCCPAC}, $i\in \mathcal{I}$ is the index of the current information bit being decoded, $j\in \mathcal{H}$ is the index of this bit with respect to $\mathbf{u}_s$ according to the mapping rule of (\ref{mapping}) (here we assume that $u_{c,i}$ is not a CRC bit in $\mathbf{c}_s$ or $\mathbf{c}_c$).

Note that $-\ln(\mathrm{P}(u_c^{[i]}|\mathbf{y}))$ is just the path metric used in channel SCL decoding\footnote{For list decoding of PAC codes, the path metric can be similarly calculated by taking the convolutional transform into account.} \cite{BS2015LLRSCL}, which can be efficiently calculated as
\begin{align}
	\mathrm{PM}_c(i)&\triangleq -\ln(\mathrm{P}(u_c^{[i]}|\mathbf{y}))\nonumber\\
	&=\sum_{j=1}^{i}\ln\big{(} 1+e^{-(1-2\hat{u}_{c,j})\cdot L_c(j)} \big{)}, \label{PMchannel}
\end{align}
where
$L_c(j)=\ln\frac{P(\mathbf{y},u_c^{[j-1]}|u_{c,j}=0)}{P(\mathbf{y},u_c^{[j-1]}|u_{c,j}=1)}$. However, $\mathrm{P}(u_s^{[j]\cap \mathcal{H}}|\mathcal{C}_s)$ (written as $\mathrm{P}(u_s^{[j]\cap \mathcal{H}})$ for short in the rest of the paper) cannot be similarly calculated because $u_s^{[j]\cap \mathcal{H}}$ does not contain the abandoned bits $u_s^{[j]\cap \mathcal{H}^C}$. According to our mapping rule and the law of total probability, 
\begin{align}
	\mathrm{P}(u_s^{[j]\cap \mathcal{H}})=\sum_{u_s^{[j]\cap\mathcal{H}^C}} \mathrm{P}(u_s^{[j]}).\label{JSCPDRule}
\end{align}
$\mathrm{P}(u_s^{[j]})$ in (\ref{JSCPDRule}) can be efficiently calculated using a similar expression to (\ref{PMchannel}) (by deleting $\mathbf{y}$ in the expressions). However, when $|[j]\cap\mathcal{H}^C|$ is large, this problem is still computationally infeasible.

Our solution to this problem is to use a source SCL decoder to approximate (\ref{JSCPDRule}). Instead of considering all possible $u_s^{[j]\cap\mathcal{H}^C}$, we only consider the ones in the source decoding list, as the candidates in the list usually are the ones with the highest probabilities. Ignoring those that are not in the list has only a little impact on the estimation, provided that the list size is large enough.

\begin{figure}[t]
	\centering
	\includegraphics[width=0.86\columnwidth]{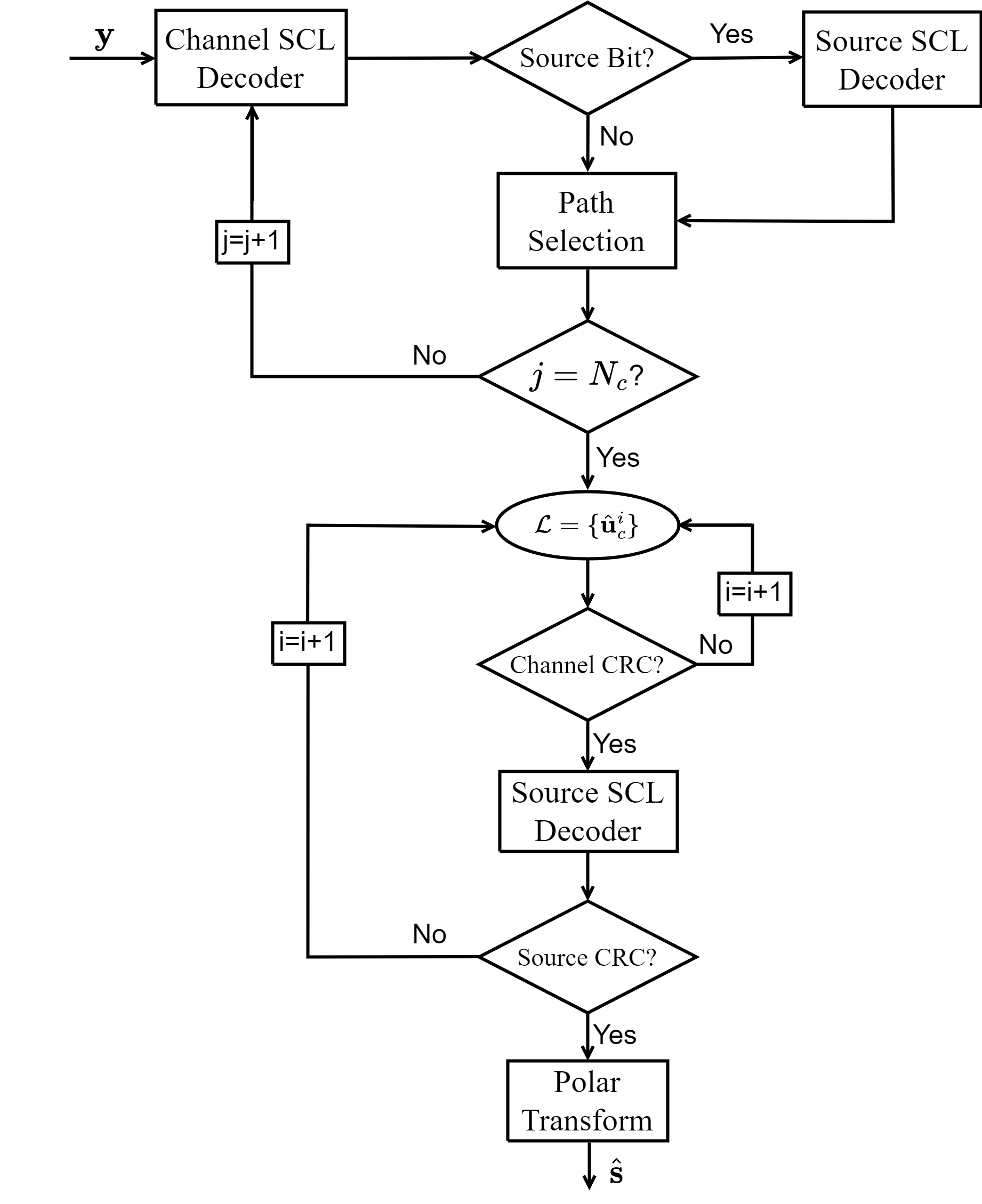}
	\caption{The joint source-channel polar/PAC decoding scheme.} 
	\label{fig:FJD}
\end{figure}

Fig. \ref{fig:FJD} shows the flowchart of the joint decoding scheme. At first, a channel list decoder decodes $\mathbf{u}_{c}$ with list size $L_c$. The list of candidates is denoted as $\mathcal{L}$. When decoding a non-CRC information bit, a source list decoder with list size $L_{sc}$ joins the decoding process to estimate the source probability $\mathrm{P}(u_s^{[j]\cap \mathcal{H}})$ for each candidate in $\mathcal{L}$. Specifically, when decoding $u_{c,i}$ ($i\in\mathcal{I}$ and $u_{c,i}$ is not a CRC bit), the path metric used for path selection consists of a channel path metric, denoted as $PM_c^{(l_c)}(i,d)$ with $l_c$ being the index of the candidate in $\mathcal{L}$ and $d\in\{0,1\}$ indicates $u_{c,i}=0$ or 1, and a source path metric, denoted as $PM_{sc}^{(l_c)}(i,d)$. $PM_c^{(l_c)}(i,d)$ is calculated using (\ref{PMchannel}), while $PM_{sc}^{(l_c)}(i,d)$ is calculated as follows
\begin{align*}
	PM_{sc}^{(l_c)}(i,d)&\triangleq -\ln\Big{[}\sum_{l_{sc}\in[L_{sc}]} \exp \big{(} -PM_s^{(l_c,l_{sc})}(j,d) \big{)} \Big{]}, \label{JSCPsource}
\end{align*}
where $PM_s^{(l_c,l_{sc})}(j,d)$ is the path metric of the $l_{sc}$-th candidate in the source decoding list for the $l_c$-th candidate in the channel decoding list.
Then path metrics are calculated as
\begin{align}
	PM^{(l_c)}(i,d)&=PM_c^{(l_c)}(i,d)+PM_{sc}^{(l_c)}(j,d)
\end{align}
and used for path selection.

After the channel list decoder has generated the final list of candidates, the following procedures are performed to recover the original source:
\begin{itemize}
	\item Sort the candidates according to their path metrics.
	\item Starting from the one with the highest probability, check whether it can pass the channel CRC.
	\item If yes, perform CA-PAC source list decoding with list size $L_s$ and check whether there exists a reconstruction result that can pass the source CRC.
	\item If yes, return the result. Otherwise try the next candidate in the channel decoder's list.
\end{itemize}

\subsection{Simulation Results}
\label{Sec:Simulation}

We consider transmitting a $\text{Bern}(0.11)$ source over a BI-AWGN channel. The channel code length is $N_c=128$ while the source length is also $N_s=128$. As benchmarks, the SSCC finite-length bound is plotted according to \cite{Kostina2012lossy} and \cite{Polyanskiy2010finite} by optimizing the source coding rate at different SNRs and the JSCC finite-length bound is plotted according to \cite{Kostina2013JSCC}. 

\begin{figure}[t]
 	\centering 
 	\includegraphics[width=0.98\columnwidth]{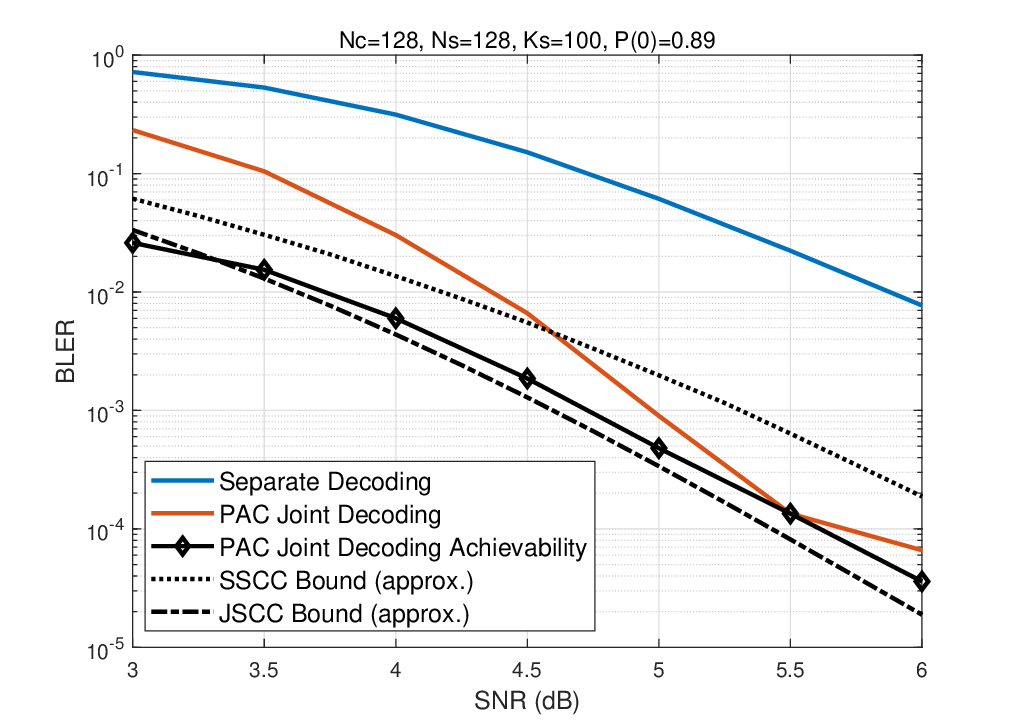}
 	\caption{Performance of the proposed joint source-channel PAC coding scheme. } 
 	\label{fig:JSCCPAC128new}
 \end{figure}

The result is shown in Fig. \ref{fig:JSCCPAC128new}. In this example, the source PAC encoder compresses a 128-bit source sequence into 100 bits, including 8 CRC bits. The channel PAC code does not use CRC, as we find that this setting optimizes the overall performance for this example. The channel decoding list size is $L_c=128$, the source decoding list size during the channel decoding process is $L_{sc}=32$, and the final source decoding list size is $L_s=128$. It can be seen that the proposed joint decoding scheme outperforms separate decoding schemes significantly. Besides, the joint decoding scheme has broken though the SSCC bound in the high SNR region and approached the JSCC bound. Part of the reason for the poor performance of separate decoding scheme is that we do not use CRC in the channel coding part. While this setting is better suited for joint decoding, it is not optimal for separate decoding. If we optimize the compression rate and allocation of source and channel CRC bits, the separate decoding scheme may approach the SSCC bound, but that is the ultimate limit.

Note that the SSCC and JSCC bounds are not achieved by a single pair of source and channel codes. At different SNRs, the optimal compression rate varies for both SSCC and JSCC. To see how close we can get to these bounds, we also optimized the compression rate for each SNR and plotted the lowest achievable BLER of the proposed scheme in the considered SNR region in Fig. \ref{fig:JSCCPAC128new}, as shown by the curve termed PAC Joint Decoding Achievability. It can be seen that this curve is very close to the JSCC bound for the whole SNR region, with a gap of about 0.2 dB at $BLER=10^{-4}$.

\section{Discussion}
\label{Sec:Conc}
In this paper, we showed that PAC codes are also finite-length bound-approaching in source coding and joint source-channel coding, just like in channel coding. The main drawback of the proposed scheme is the decoding complexity, as we use a list source decoder to estimate the source probability for each candidate in the channel decoding list. Nevertheless, we successfully showed that the JSCC finite-length bound can actually be approached with practical codes. How to reduce the complexity while not sacrificing too much performance is worth future research. 

\bibliographystyle{IEEEtran}
\bibliography{JSCC_WCSP}

\end{document}